# Tuning of Carrier Concentration and Superconductivity in High-Entropy-Alloy-Type Metal Telluride $(AgSnPbBi)_{(1-x)/4}In_xTe$

Md. Riad Kasem[1], Ryota Ishii[2], Takayoshi Katase[3], Osuke Miura[2], Yoshikazu Mizuguchi[1]*

1. Department of Physics, Tokyo Metropolitan University, 1-1, Minami-osawa, Hachioji, 192-0397, Japan.
2. Department of Electrical Engineering and Computer Science, Tokyo Metropolitan University, 1-1, Minami-osawa, Hachioji, 192-0397, Japan.
3. Laboratory for Materials and Structures, Institute of Innovative Research, Tokyo Institute of Technology, 4259 Nagatsuta, Midori, Yokohama, 226-8503, Japan.

Corresponding author: Yoshikazu Mizuguchi (mizugu@tmu.ac.jp)

**Abstract**

High-entropy-alloy-type (HEA-type) compound superconductors have been drawing much attention as a new class of exotic superconductors with local structural inhomogeneity. NaCl-type (Ag,In,Sn,Pb,Bi)Te is a typical HEA-type superconductor, but the carrier doping mechanism had been unclear. In this study, we synthesized (Ag,In,Sn,Pb,Bi)Te with various In concentration using high-pressure synthesis: the studied system is $(AgSnPbBi)_{(1-x)/4}In_xTe$ ($x$ = 0–0.4). Single-phase samples were obtained for $x$ = 0–0.3. A semiconductor-like temperature dependence of resistivity was observed for $x$ = 0, while superconductivity appeared for the In-doped samples. The highest transition temperature ($T_c$) was 3.0 K for $x$ = 0.3. The Seebeck coefficient decreases with increase of $x$, which suggests that $In^{3+}$ generates electron carriers in $(AgSnPbBi)_{(1-x)/4}In_xTe$. Tuning of carrier concentration and superconducting properties of (Ag,In,Sn,Pb,Bi)Te would be useful for further investigation of exotic superconductivity in the HEA-type compound.

NaCl-type metal tellurides ($M$Te) have been extensively studied owing to its characteristics as a superconductor [1–5], a thermoelectric material [6–9], and a topological-crystalline insulator [10–12]. In addition, carrier-doped SnTe has been studied as a candidate system of topological superconductor [13]. Recently, high-entropy-alloy-type (HEA-type) metal tellurides (or metal chalcogenides) have been developed as a high-performance thermoelectric material [14–16] and a new class of exotic superconductor with highly inhomogeneous local structures [17–19]. HEAs are alloys containing five or more elements with a solution ratio of 5–35 at% [20–22] and have high configurational entropy of mixing ($\Delta S_{mix}$), which is given by $\Delta S_{mix} = -R \sum^{i} c_i \ln c_i$ where $R$ and $c_i$ are gas constant and atomic fraction, respectively. Due to high $\Delta S_{mix}$, several functionalities have been improved. On superconductivity, the first HEA superconductor was discovered in 2014 in a Ti-Zr-Hf-Nb-Ta alloy [23]. Various HEA superconductors have been synthesized, and the similarity and the differences between conventional alloys, amorphous, and HEA have been studied [24,25]. One of the notable features of the HEA superconductor Ti-Zr-Hf-Nb-Ta is the robustness of superconductivity under extremely high pressures [26]. To develop exotic superconductors with HEA states, we have applied the concept of HEA to compound superconductors with two or more crystallographic sites: for example, NaCl-type $M$Te, layered BiS$_2$-based, CuAl$_2$-type $Tr$Zr$_2$ ($Tr$: transition metal), and A15-type Nb$_3X$ (X = Al, Si, Ga, Ge, Sn) [17, 27–30]. Very recently, the robustness of superconductivity under high pressures has been observed in HEA-type $M$Te, (Ag,In,Sn,Pb,Bi)Te [31], while the transition temperature ($T_c$) of PbTe ($\Delta S_{mix} = 0$) decreases with pressure. Since the robustness of superconductivity under high pressures has been commonly observed in alloy-based Ti-Zr-Hf-Nb-Ta and compound-based (Ag,In,Sn,Pb,Bi)Te, the phenomenon would be universal among HEA-type superconductors. Therefore, further understanding of the superconducting and physical properties of the (Ag,In,Sn,Pb,Bi)Te system is important. Particularly, the carrier doping mechanism has not been clarified.

Basically, $M$Te is composed of $M^{2+}$ and Te$^{2-}$ in valence states. For example, SnTe and PbTe are semiconductors with a band gap, while the band structure of SnTe has band inversion, and an ordinary band gap is present for PbTe. In InTe, valence skipping states of In$^+$ and In$^{3+}$ are there, and the average valence state is +2. InTe exhibits metallic conductivity and superconductivity below 3 K [3]. In our recent study on (Ag,Sn,Bi)Te [32], we found that the partial substitution of Ag$^+_{0.5}$Bi$^{3+}_{0.5}$ (average valence of +2) does not induce metallic character in (Ag,Sn,Bi)Te and proposed that band inversion is present in all $x$ in Sn$_{1-2x}$(AgBi)$_x$Te. To induce superconductivity in the (Ag,Sn,Bi)Te system, the partial substitution of In ion was essential [32]. Motivated by those facts, we herein investigate the doping effects in the HEA-type (Ag,In,Sn,Pb,Bi)Te system in a compositional formula of $(AgSnPbBi)_{(1-x)/4}In_xTe$ where $x$ corresponds to the nominal In concentration. We found that the sample with $x = 0$ exhibits semiconducting characteristics. According to theoretical calculation on lattice constant and band structure for a SnTe-based structure [32], $Ag_{0.25}Sn_{0.25}Pb_{0.25}Bi_{0.25}Te$ ($x = 0$) is expected to possess band inversion and thus to be a topological insulator. Superconductivity was induced by the In substitution in $(AgSnPbBi)_{(1-x)/4}In_xTe$. The highest $T_c$ was observed for $x = 0.3$, and the increase of In concentration increased electron carriers in the system.

The polycrystalline samples of $(AgSnPbBi)_{1-4x/4}In_xTe$ with nominal $x$ of 0, 0.1, 0.2, 0.3, and 0.4 were prepared by solid-state reaction. To obtain precursor powders, the powders of Ag (99.9%) and grains of Sn (99.99%), Pb (99.99%), Bi (99.99%), In (99.99%), and Te (99.99%) with the nominal compositions were mixed and sealed in an evacuated quartz tube. The precursor powders were obtained by melting the samples in a furnace at 800 °C for 15 hours, followed by furnace cooling. The obtained precursors were powdered, palletized into 5 mm diameter, and inserted in a high-pressure-synthesis cell. We used a cubic-anvil-type 180-ton press and annealed the samples under 3 GPa at 500 °C for 30 min. Powder X-ray diffraction (XRD) was performed on a Miniflex600 (RIKAKU) diffractometer equipped with a CuK$_{α1}$ radiation ($λ = 1.54$ Å) by the $θ$-$2θ$ method. Obtained XRD patterns were refined by Rietveld analysis using RIETAN-FP software [33] to estimate lattice constant. The crystal structure image was drawn using VESTA software [34]. The actual compositions of the obtained samples were analyzed by energy-dispersive X-ray spectroscopy (EDX) with SwiftED analyzer (Oxford) on a scanning electron microscope TM-3030 (Hitachi Hightech). Temperature dependence of electrical resistivity was measured by the standard four-probe method on a GM refrigerator system (AXIS). Ag paste was used to attach Au wire (25 μm in diameter) on the surface of the samples. The temperature dependence of magnetic susceptibility was measured under applied field of 10 Oe [zero-field-cooling (ZFC) and field-cooling (FC) modes] on a superconducting quantum interference device (MPMS3, Quantum Design). Seebeck coefficient ($S$) at room temperature was measured under steady-state, where the thermo-electromotiveforce ($ΔV$) and the temperature difference ($ΔT$) were simultaneously measured, and the $S$ was determined from the slope of $ΔV/ΔT$.

In Table 1, the nominal and actual (EDX) compositions are summarized. The averaged actual compositions are close to the nominal values. Thus, we call samples with nominal $x$ in this paper.

Figures 1(a) and 1(b) show normalized powder XRD patterns for the $(AgSnPbBi)_{(1-x)/4}In_xTe$ ($x = 0, 0.1, 0.2\ 0.3$, and 0.4) samples after high-pressure annealing.

The peaks could be indexed by a NaCl-type model [space group: $Fm$-$3m$, Fig. 1(d)] for all samples. For $x$ = 0.0–0.3, single-phase samples were obtained. For $x$ = 0.4, the XRD peaks are broadened due to the presence of the NaCl-type impurity phases with a larger (peak A) and smaller (peak B) lattice constant. The lattice constants roughly estimated from the positions of peak A and B suggest the presence of SnTe and $AgInTe_2$ [35]. Therefore, the solubility limit of In in $(AgSnPbBi)_{(1-x)/4}In_xTe$ is $x$ = 0.3–0.4. Figure 1(c) shows the nominal $x$ dependence of lattice constant of $x$ = 0–0.4 and InTe ($x$ = 1) [5]. For $x$ = 0, 0.1, 0.2, 0.3, and 1, data points obey the Vegard's law. The reason why the data point for $x$ = 0.4 deviates from the Vegard's-law line would be the presence of impurity phases mentioned above.

Figure 2(a) displays the temperature dependence of resistivity ($\rho$-$T$) for $x$ = 0. The $\rho$ increases on cooling at low temperatures, which suggests the presence of a band gap for $x$ = 0. In our earlier work on (Ag,Sn,Bi)Te, it was found that the $Ag_{0.5}Bi_{0.5}$ substitution for the Sn site does not generate carriers, which is consistent with the present result. Therefore, regarding $x$ = 0 ($Ag_{0.25}Sn_{0.25}Pb_{0.25}Bi_{0.25}Te$) as a non-doped system is reasonable to examine the carrier doping effects in $(AgSnPbBi)_{(1-x)/4}In_xTe$. Figure 2(b) shows the $\rho$-$T$ for $x$ = 0.1, 0.2, 0.3, and 0.4. The resistivity for $x$ = 0.1 is clearly lower than that for $x$ = 0. Although the $\rho$-$T$ for $x$ = 0.1 still exhibits a semiconducting-trend, metallic trend of $\rho$-$T$ is seen for $x$ = 0.2, 0.3, and 0.4. On the basis of those results, we consider that the partial substitution of In for the $M$ site generates carriers and make the system metallic. The low-temperature $\rho$-$T$ for $x$ = 0.1, 0.2, 0.3, and 0.4 is displayed in Fig. 2(c). The highest resistive $T_c$ is observed for $x$ = 0.3. The $T_c$s estimated from resistivity onset ($T_c^{\rho(\text{onset})}$) and zero resistivity ($T_c^{\rho(\text{zero})}$) are listed in Table 1, and the $T_c^{\rho(\text{zero})}$ is plotted in Fig. 2(d) as a function of nominal $x$. The highest $T_c^{\rho(\text{zero})}$ is 2.8 K for $x$ = 0.3.

Figure 3 shows the temperature dependence of magnetization for $x$ = 0.2, 0.3, 0.4 in the main panel and $x$ = 0 and 0.1 in the inset. For $x$ = 0.2, 0.3, 0.4, large shielding signals are observed, which indicates the emergence of bulk superconductivity. For $x$ = 0.1, the observed shielding fraction is quite small, suggesting filamentary superconductivity. $T_c$ estimated from magnetization transition ($T_c^{\text{mag}}$) is listed in table 1. The highest $T_c^{\text{mag}}$ = 2.7 K was observed for $x$ = 0.3 and 0.4.

To discuss about the characteristics of the carriers generated by In ion substitution in $(AgSnPbBi)_{(1-x)/4}In_xTe$, $S$ was measured at room temperature. In Fig. 4(a), the $\Delta V/\Delta T$ plots for all the samples are shown, which exhibits a linear relationship. From the slope of $\Delta V/\Delta T$, $S$ was calculated and plotted in Fig. 4(b) as a function of nominal $x$. For $x$ = 0, $S$ is -13.6 µV/K suggesting that the phase is an $n$-type semiconductor. With increasing $x$, the absolute value of $S$ decreases and remains negative for $x$ = 0.1, 0.2, 0.3.

Although the signal of $S$ for $x = 0.4$ is positive, we cannot quantitatively discuss about the result because of the presence of NaCl-type impurity phases in the sample. Therefore, the In ion substitution basically generates electron carriers in $(AgSnPbBi)_{(1-x)/4}In_xTe$ at least for $x = 0$–$0.3$.

Here, we briefly discuss about the valence states of In doped in various NaCl-type $M$Te compounds. In pure InTe, $In^+$ and $In^{3+}$ are present, and metallic and superconducting properties are observed [3]. This trend is confirmed by band calculation [3,5]. In the case of In-doped PbTe [5], $S$ is positive for a wide range of In concentrations, which suggests that the doped In is $In^+$ and providing hole carriers. For In-doped SnTe [1], Hall coefficient suggests switching of dominant carriers from holes to electrons. In In-doped (Ag,Sn,Bi)Te, In doping generates electron carriers as in the case of $(AgSnPbBi)_{(1-x)/4}In_xTe$. Therefore, in (Ag,Sn,Bi)Te and (Ag,Sn,Pb,Bi)Te, the valence state of In is $In^{3+}$. These trends can be understood by the difference in lattice constant. PbTe has the largest lattice constant, and hence, $In^+$ is preferred when being doped for the Pb site because ionic radius of $In^+$ is larger than that of $In^{3+}$. In contrast, the lattice constant for (Ag,Sn,Bi)Te and (Ag,Sn,Pb,Bi)Te is smaller than that of PbTe. In such a small unit cell, the form of of $In^{3+}$ is preferred. The trends discussed here show that the In valence states in a $M$Te structure can be modified by tuning the unit-cell size. The information will be useful for further development of $M$Te superconductors and related functional materials. To clarify the details of valence states of In in $M$Te, high-resolution spectroscopy is desired.

In conclusion, we synthesized polycrystalline samples of $(AgSnPbBi)_{(1-x)/4}In_xTe$ using high-pressure annealing. Single-phase samples were obtained for $x = 0, 0.1, 0.2, 0.3$. The $x = 0$ ($Ag_{0.25}Sn_{0.25}Pb_{0.25}Bi_{0.25}Te$) sample showed a semiconducting behavior in the resistivity measurement. Through resistivity and magnetization measurements, superconducting properties were investigated for the obtained samples, and the highest $T_c$ was observed for $x = 0.3$. In addition, Seebeck coefficient measurements revealed that the $In^{3+}$ substitution generates electron carriers in $(AgSnPbBi)_{(1-x)/4}In_xTe$, which is essential for metallic conductivity and superconductivity in the HEA-type $M$Te system. Furthermore, we proposed that the unit-cell size of $M$Te affects the valence state of In doped in a NaCl-type structure.

**Acknowledgements**
This work was partly supported by grants in Aid for Scientific Research (KAKENHI) (18KK0076, 21K18834, and 21H00151) and the Tokyo Government Advanced Research (H31-1). A part of this study was supported by the Collaborative Research Project of Laboratory for Materials and Structures, Institute of Innovative Research, Tokyo Institute of Technology.

**References**

1.  N. Haldolaarachchige, Q. Gibson, W. Xie, M. B. Nielsen, S. Kushwaha, and R. J. Cava, Phys. Rev. B 93, 024520 (2016).
2.  A. S. Erickson, J.-H. Chu, M. F. Toney, T. H. Geballe, and I. R. Fisher, Phys. Rev. B 79, 024520 (2009).
3.  K. Kobayashi, Y. Ai, H. O. Jeschke, J. Akimitsu, Phys. Rev. B 97, 104511 (2018).
4.  Y. Mizuguchi and O. Miura, J. Phys. Soc. Jpn. 85, 053702 (2016).
5.  M. Katsuno, R. Jha, K. Hoshi, R. Sogabe, Y. Goto and Y. Mizuguchi, Condens. Matter 5, 14 (2020).
6.  J. P. Heremans, V. Jovovic, E. S. Toberer, A. Saramat, K. Kurosaki, A. Charoenphakdee, S. Yamanaka, and G. Jeffrey Snyder, Science 321, 554 (2008).
7.  G. Tan, S. Hao, R. C. Hanus, X. Zhang, S. Anand, T. P. Bailey, A. J. Rettie, X. Su, C. Uher, V. P. Dravid, ACS Energy Letters, 3, 705 (2018).
8.  Y. Zhong, J. Tang, H. Liu, Z. Chen, L. Lin, D. Ren, B. Liu, and R. Ang, ACS Appl. Mater. Interfaces 12, 49323 (2020).
9.  Q. Zhang, B. Liao, Y. Lan, K. Lukas, W. Liu, K. Esfarjani, C. Opeil, D. Broido, G. Chen, and Z. Ren, Proc. Natl. Acad. Sci. USA. 110, 13261 (2013).
10. T. H. Hsieh, H. Lin, J. Liu, W. Duan, A. Bansil, and L. Fu, Nat. Commun. 3, 982 (2012).
11. Y. Tanaka, Z. Ren, T. Sato, K. Nakayama, S. Souma, T. Takahashi, K. Segawa, and Y. Ando, Nat. Phys. 8, 800 (2012).
12. Y. Ando, J. Phys. Soc. Jpn. 82, 102001 (2013).
13. S. Sasaki, Z. Ren, A. A. Taskin, K. Segawa, L. Fu, and Y. Ando. Phys. Rev. Lett. 109, 217004 (2012).
14. B. Jiang, Y. Yu, J. Cui, X. Liu, L. Xie, J. Liao, Q. Zhang, Y. Huang, S. Ning, B. Jia, B. Zhu, S. Bai, L. Chen, S. J. Pennycook, and J. He, Science 371, 830 (2021).
15. B. Jiang, Y. Yu, H. Chen, J. Cui, X. Liu, L. Xie, and J. He, Nat. Commun. 12, 3234 (2021).
16. A. Yamashita, Y. Goto, C. Moriyoshi, Y. Kuroiwa, and Y. Mizuguchi, Mater. Res. Lett. 9, 366 (2021).
17. Y. Mizuguchi, J. Phys. Soc. Jpn. 88, 124708 (2019).
18. M. R. Kasem, K. Hoshi, R. Jha, M. Katsuno, A. Yamashita, T. D. Goto, Matsuda, Y. Aoki, Y. Mizuguchi, Appl. Phys. Express 13, 033001 (2020).
19. A. Yamashita, R. Jha, Y. Goto, T. D. Matsuda, Y. Aoki, and Y. Mizuguchi Dalton Trans. 49, 9118 (2020).

20. J. W. Yeh, S. K. Chen, S. J. Lin, J. Y. Gan, T. S. Chin, T. T. Shun, C. H. Tsau, and S. Y. Chang, Adv. Eng. Mater. 6, 299 (2004).
21. M. H. Tsai and J. W. Yeh, Mater. Res. Lett. 2, 107. (2014).
22. H. Inui, K. Kishida, and S. Z.H. Chen, Mater. Trans. 63, 394 (2022).
23. P. Koželj, S. Vrtnik, A. Jelen, S. Jazbec, Z. Jagličić, S. Maiti, M. Feuerbacher, W. Steurer, and J. Dolinšek, Phys. Rev. Lett. 113, 107001 (2014).
24. L. Sun and R. J. Cava, Phys. Rev. Mater. 3, 090301 (2019).
25. J. Kitagawa, S. Hamamoto, and N. Ishizu, Metals 10, 1078 (2020).
26. J. Guo, H. Wang, F. von Rohr, Z. Wang, S. Cai, Y. Zhou, K. Yang, A. Li, S. Jiang, Q. Wu, R. J. Cava, and L. Sun, Proc. Natl. Acad. Sci. U.S.A. 114, 13144 (2017).
27. R. Sogabe, Y. Goto, and Y. Mizuguchi, Appl. Phys. Express 11, 053102 (2018).
28. Y. Mizuguchi, M. R. Kasem, T. D. Matsuda, Mater. Res. Lett. 9, 141 (2020).
29. A. Yamashita, T. D. Matsuda, and Y. Mizuguchi, J. Alloys Compd. 868, 159233 (2021).
30. Y. Mizuguchi and A. Yamashita, IntechOpen (online book chapter), DOI: 10.5772/intechopen.96156.
31. T. Mitobe, K. Hoshi, M. R. Kasem, R. Kiyama, A. Yamashita, R. Higashinaka, T. D. Matsuda, Y. Aoki, Y. Goto and Y. Mizuguchi Sci. Rep. 11, 22885 (2021).
32. M. R. Kasem, Y. Nakahira, H. Yamaoka, R. Matsumoto, A. Yamashita, H. Ishii, N. Hiraoka, Y. Takano, Y. Goto, Y. Mizuguchi, arXiv:2112.06461 (2021).
33. F. Izumi and K. Momma, Solid State Phenom. 130, 15 (2007).
34. K. Momma and F. Izumi, J. Appl. Cryst. 41, 653 (2008).
35. T. Bovornratanaraks, K. Kotmool, K. Yoodee, M. I. McMahon, and D. Ruffolo, J. Phys.: Conf. Ser. 215, 012008 (2010).

Table 1. Actual *M*-site composition estimated by EDX, $\Delta S_{mix}/R$ at the *M* site, $T_c$s estimated from magnetization ($T_c^{mag}$), resistivity onset ($T_c^{\rho(onset)}$), and zero resistivity ($T_c^{\rho(zero)}$) for $(AgSnPbBi)_{(1-x)/4}In_xTe$.

| Nominal *x* | Actual *M*-site composition | $\Delta S_{mix}/R$ | $T_c^{mag}$ (K) | $T_c^{\rho(onset)}$ (K) | $T_c^{\rho(zero)}$ (K) |
|---|---|---|---|---|---|
| 0 | $Ag_{0.27}Sn_{0.26}Pb_{0.22}Bi_{0.25}$ | 1.38 | - | - | - |
| 0.1 | $Ag_{0.24}Sn_{0.23}Pb_{0.21}Bi_{0.22}In_{0.10}$ | 1.57 | 2.0 | 2.0 | 1.6 |
| 0.2 | $Ag_{0.20}Sn_{0.22}Pb_{0.19}Bi_{0.19}In_{0.20}$ | 1.61 | 2.4 | 2.8 | 2.6 |
| 0.3 | $Ag_{0.17}Sn_{0.19}Pb_{0.15}Bi_{0.18}In_{0.31}$ | 1.57 | 2.7 | 3.0 | 2.8 |
| 0.4 | $Ag_{0.15}Sn_{0.16}Pb_{0.15}Bi_{0.14}In_{0.40}$ | 1.51 | 2.7 | 3.0 | 2.6 |

Figures

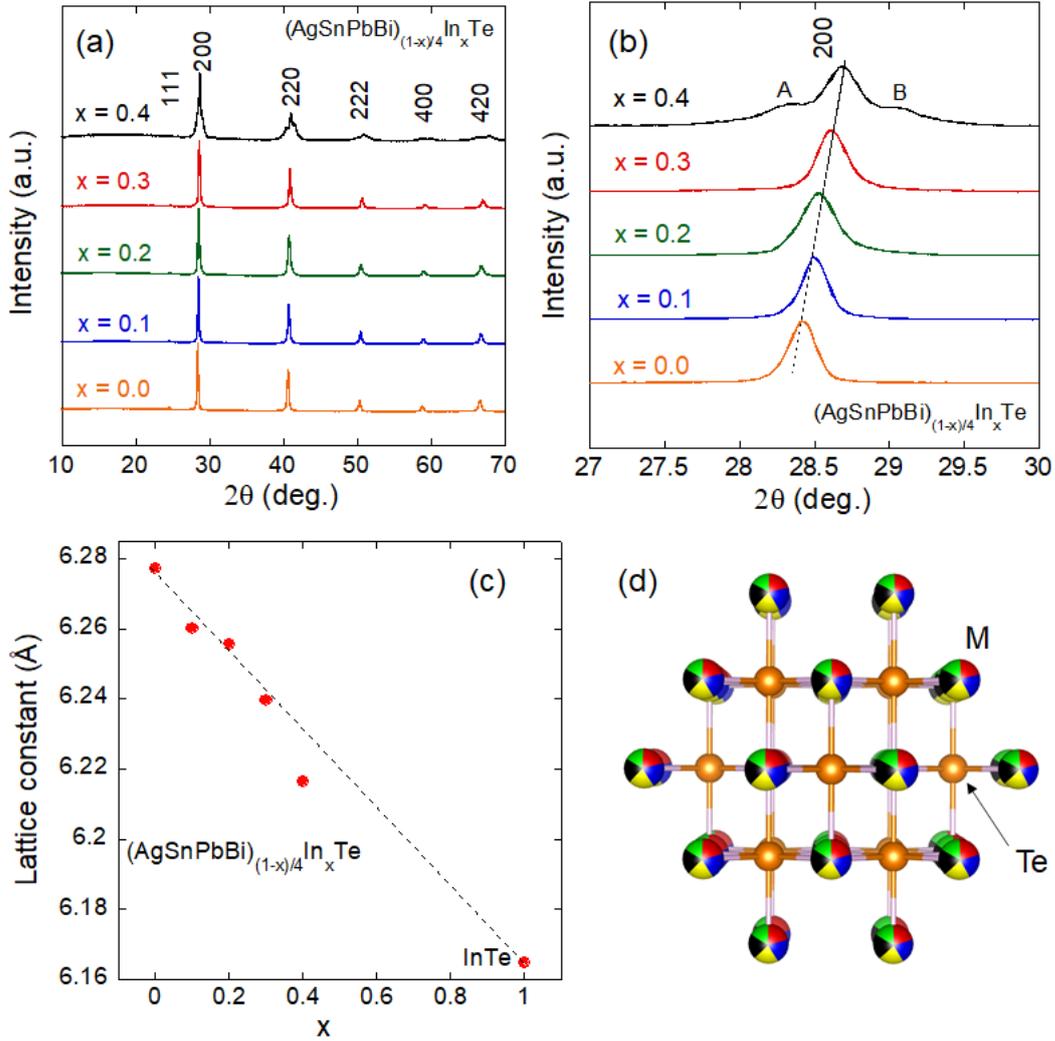

Fig. 1. (Color online) (a) Powder XRD patterns of (AgSnPbBi)$_{(1-x)/4}$In$_x$Te. The numbers in the figure are Miller's indices. (b) Zoomed XRD patterns near the 2 0 0 peak. Peak A and B are impurity phases of NaCl-type $M$Te with a larger and smaller lattice constant, respectively. (c) Nominal $x$ dependence of lattice constant. The dashed linear line is eye-guide. (d) Schematic image of the crystal structure of HEA-type $M$Te.

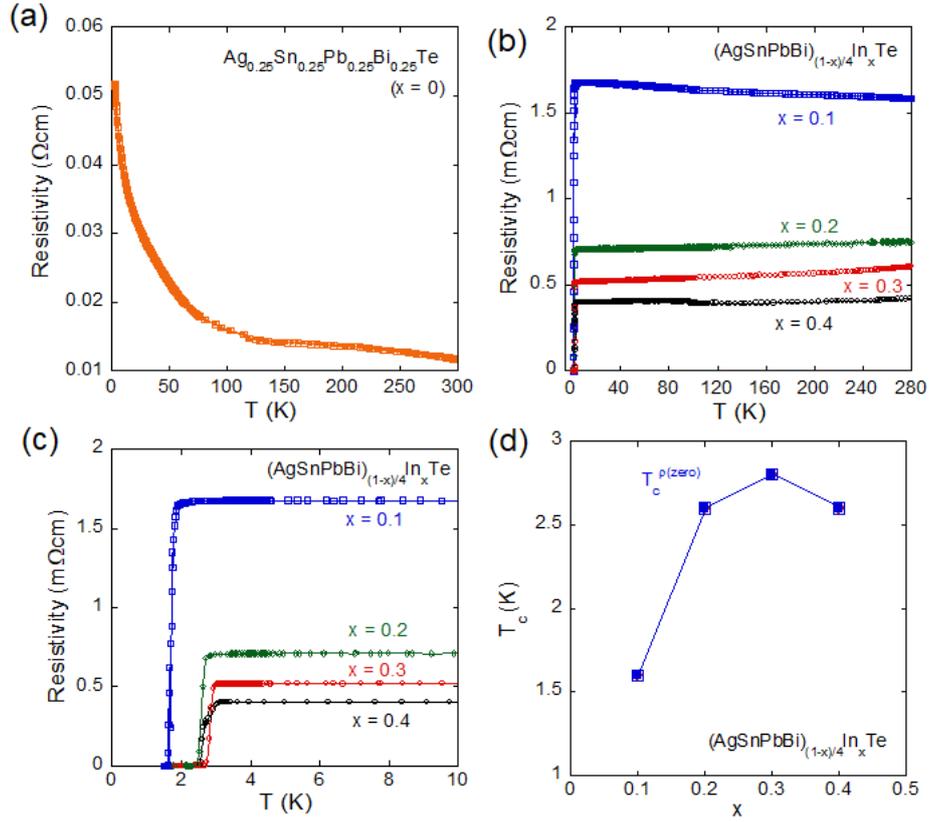

Fig. 2. (Color online) (a) Temperature dependence of resistivity ($\rho$-$T$) for $x$ = 0. (b, c) $\rho$-$T$ for x = 0.1, 0.2, 0.3, and 0.4. (d) Nominal $x$ dependence of $T_c^{\rho(\text{zero})}$.

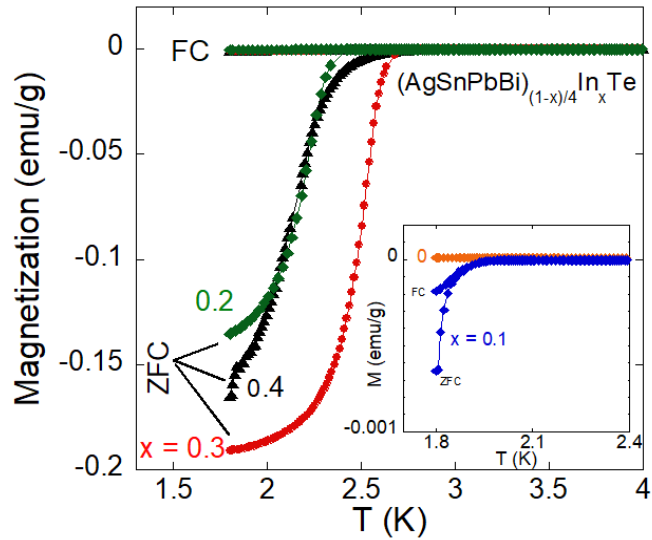

Fig. 3. (Color online) Temperature dependences of magnetization ($H$ = 10 Oe) for $x$ = 0–0.4 of (AgSnPbBi)$_{(1-x)/4}$In$_x$Te.

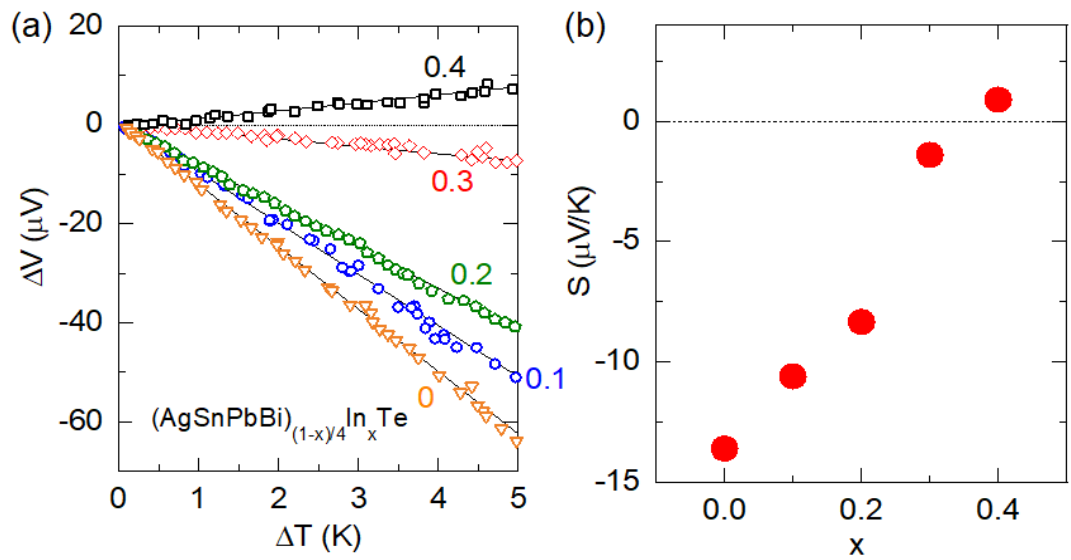

Fig. 4. (Color online) (a) $\Delta V$-$\Delta T$ plots at room temperature for $x$ = 0–0.4 of $(AgSnPbBi)_{(1-x)/4}In_xTe$. (b) Nominal $x$ dependence of Seebeck coefficient at room temperature.